\documentclass[traditabstract]{aa} % for the abstract without structuration 
\usepackage{graphicx}
\usepackage{txfonts}
\usepackage{lscape}
\usepackage{amssymb}
\begin{document}
\newcommand{\mincir}{\raise
-2.truept\hbox{\rlap{\hbox{$\sim$}}\raise5.truept\hbox{$<$}\ }}

   \title{The Activity of the Neighbours of Seyfert Galaxies.}

   \author{E.Koulouridis\inst{1},  
    M.Plionis\inst{2,3}, 
    V.Chavushyan\inst{3}, 
    D.Dultzin\inst{4}, 
    Y.Krongold\inst{4}, 
    I.Georgantopoulos\inst{1},
    J. Le\'on-Tavares\inst{5,6}}

   \institute{
    Institute of Astronomy \& Astrophysics, National Observatory of Athens,
    Palaia Penteli 152 36, Athens, Greece.
\and Physics Department of Aristotle University of Thessaloniki, University Campus, 54124, Thessaloniki, Greece
    \and Instituto Nacional de Astrof\'{\i}sica Optica y Electr\'onica, Puebla, C.P. 72840,M\'exico.
    \and Instituto de Astronom\'ia, Universidad Nacional Aut\'onoma
    de M\'exico, Apartado Postal 70-264, M\'exico, D. F. 04510, M\'exico
    \and Finnish Centre for  Astronomy with ESO (FINCA), University of Turku,
     V\"ais\"al\"antie 20, FI-21500  Piikki\"o, Finland
    \and Aalto University Mets\"ahovi Radio Observatory,  Mets\"ahovintie 114, FIN-02540
     Kylm\"al\"a, Finland}

   \date{\today}

% \abstract{}{}{}{}{} 
% 5 {} token are mandatory
 
  \abstract{
  We present a follow-up study on a series of papers concerning the role
of close interactions as a possible triggering mechanism of
AGN activity. We have already studied the close ($\leq
100 \; h^{-1}$kpc) and the large scale ($\leq$ 1 $h^{-1}$ Mpc)
environment of a local sample of Sy1, Sy2 and bright IRAS galaxies (BIRG) and their
respective control samples. The results led us to the conclusion that a close encounter appears
capable of activating a sequence where an absorption line galaxy (ALG) galaxy becomes first a
starburst, then a Sy2 and finally a Sy1. 
Here we investigate the activity of neighboring galaxies of different
types of AGN, since both galaxies of an interacting pair should be
affected.
To this end we present the optical spectroscopy and X-ray imaging of 30 neighbouring galaxies
around two local ($z\mincir 0.034$) samples of 10 Sy1 and 13 Sy2 galaxies. Although this is a pilot study of a small sample,
various interesting trends have been discovered
implying physical mechanisms which may lead to different Seyfert
types.
Based on the optical spectroscopy we find that more than
70\% of all neighbouring galaxies exhibit star forming and/or nuclear
activity (namely recent star formation and/or AGN), while an
additional X-ray analysis showed that this percentage might be
significantly higher. Furthermore, we find a statistically significant
correlation, at a 99.9\% level, between the value of the
neighbour's [OIII]/H$\beta$ ratio and the activity type of
the central active galaxy, i.e. the neighbours of Sy2 galaxies are
systematically more ionized than the neighbours of Sy1s. This result,
in combination with trends found using the
Equivalent Width of the H$\alpha$ emission line and the
stellar population synthesis code STARLIGHT, indicate differences
in the stellar mass, metallicity and star formation history between
the samples. Our results point towards a link between close galaxy
interactions and activity and also provide more clues regarding the
possible evolutionary sequence inferred by our previous studies.}

   \keywords{Galaxies: Active, Galaxies: Seyfert, Galaxies:
     interactions, Galaxies: nuclei, X-rays: Galaxies, Cosmology:
     Large-Scale Structure of Universe}
\authorrunning{Koulouridis et al.}
\titlerunning{Neighbours of Seyfert Galaxies.}
   \maketitle
%
%________________________________________________________________
\section{Introduction}

%Since the discovery of Active Galactic Nuclei (AGN), 
%significant effort has been put in the attempt to reveal their
%nature. However, the lack of detailed knowledge of key aspects of the AGN mechanism
%leaves us with many scattered pieces of information. Theory is
%unable in most cases to explain observations and observations still cannot clearly 
%resolve the galactic nuclei in order to confirm
%theories. Radio-loud and radio-quiet AGN, QSO type I and II, Sy1
%and Sy2 galaxies, LINERs, transition galaxies between different states
%(TOs) and starburst galaxies are some of the pieces of the puzzle
%that we are called to unify. Examination of 
The properties of the host
galaxies of the different types of AGN and their environment, up to
several hundred kpc, can give us valuable information on the nature of the general AGN
population, as well as on different properties of each AGN subtype. 
In addition, the availability nowadays of large automatically
constructed galaxy catalogues, like the SDSS, can provide 
the necessary statistical significance for these type of analyses. 
However, great caution should be taken when interpreting
results based on large databases, since the larger the sample size the
less control usually one has on the spectral and other details of the individual galaxy entries.
It could then be difficult to address important questions, such
as : Do the Unification paradigm explains all cases of type 1 and type 2 AGN?
What is the true connection between galaxy interactions, star formation and
nuclear activity? What is the lifetime of these phenomena? How do LINERs
fit in the general picture and can all be considered AGN? Do
evolutionary trends affect the AGN phenomenology?

Nowadays, it is widely accepted that the accretion of
material into a massive black hole (MBH), located at the galactic
center, is responsible for the detected excess emission (radiation not
emitted by stellar photospheres) in the AGN's spectra and such black
holes do exist in all elliptical galaxies and spiral galaxy bulges 
(Kormendy and Richstone 1995; Magorrian et al. 1998), including our
own (e.g. Melia \& Falcke 2001). However, we still lack a complete
understanding of the various aspects of activity, for example the
triggering mechanism and the feeding of the black hole, the physical
properties of the
accretion disk and the obscuring torus predicted by the unified
scheme (Antonucci et al. 1993), the origin of jets in radio loud
objects, the connection with star formation and
the role of the AGN feedback. Even the exact mechanism that
produces the observed IR, X-ray, and gamma-ray emission, is still only partially
understood (e.g. Le{\'o}n-Tavares et al. 2011). The unification model itself, although successful in many
cases, has not been able to fully explain all the AGN phenomenology (among others, the role of
interactions on induced activity; Koulouridis et al. 2006a,b and
references therein).

Despite observational difficulties and limitations, there have been
many attempts, based on different diagnostics, to investigate the
possible triggering mechanisms of nuclear activity. Most agree that
the accretion of material into a MBH (Lynden-Bell 1969) is the
mechanism responsible for the emission, but it is still necessary to
understand the feeding mechanism of the black hole. It is known and
widely accepted that interactions between galaxies can force gas and
molecular clouds towards the galactic center, where they become
compressed and produce starburst events.(e.g. Li et al. 2008;
Ellison et al. 2008; Ideue et al. 2012). Many also believe that the
same mechanism could give birth to an active nucleus 
(e.g. Umemura 1998; Kawakatu et al. 2006; Ellison et al. 2011;
Silverman et al. 2011, Villforth et al. 2012). Despite the fact that
the exact mechanism is still unknown, in the local universe a minimum
accretion rate of $\sim 10^{-6\pm1}  M_\odot$/yr is needed in order to fuel
the black hole (Ho 2008). At such low accretion rates, compared to the host galaxy, nuclear activity is probably relatively weak and most of the spectral signatures of the AGN are "buried". Theoretically the feeding of the black hole can only be achieved by means of a non axisymmetric perturbation which induces mass inflow. Such
perturbations can be provided by interactions and the result of the
inflow is the feeding of the black hole and the activation of the AGN
phase, maybe $\sim \rm 50-250\; Myr$ after the initial interaction
took place (see below). An interaction certainly predicts such a time
delay, since after the material has piled up around the inner Linblad
resonance, enhancing star formation, it can be channeled towards the
nucleus by loosing significant amounts of angular momentum, a process
which is not instantaneous.

Indeed, post starburst stellar populations have been
observed around AGN (Dultzin-Hacyan \& Benitez 1994; Maiolino \& Rieke 1995;
Nelson \& Whittle 1996; Hunt et al. 1997; Maiolino et al. 1997; Cid Fernandes,
Storchi-Bergmann \& Schmitt 1998; Boisson et al. 2000, 2004;
Cid Fernandes et al. 2001, 2004, 2005) and in close proximity to the
core ($\sim$50 pc). This fact implies the continuity of
these two states and a delay of 50-250 Myr between the onset of the
starburst and the feeding of the AGN (e.g., M{\"u}ller S{\'a}nchez et
al. 2008; Wild, Heckman, \& Charlot 2010; Davies et al. 2012), which
may reach the peak of its activity 
after $\sim$500 Myr (Kaviraj et al. 2011). Ballantyne, Everett \&
Murray (2006) studying the Cosmic XRay Background (CXRB) concluded
that Seyfert galaxies (dominating in the production of the CXRB) are
likely fueled by minor mergers or interactions that can trigger a
circumnuclear star formation event, but that there may be a
significant delay between the interaction and the ignition of the
nucleus. Davies et al. (2007), analyzing star formation in the nuclei
of nine Seyfert galaxies found recent, but no longer active,
starbursts which occurred 10 - 300 Myr ago. Further support for an
interaction-activity relation was recently provided by HI observations
of Tang et al. (2008), who found that 94\% of the Seyfert galaxies in
their sample were disturbed in contrast to their control sample (where
only 19\% were disturbed), but see also Georgakakis et al. (2009) and
Cisternas et al. (2011) in the AEGIS and cosmos surveys, respectively.

%\subsection{This work}

This paper is the third in a series of 3-dimensional studies of the
environment of active galaxies (Koulouridis et al. 2006a,b), 
extending previous 2D analyses (Dultzin et al. 1999, Krongold et
al. 2002) in an effort to shed more light to the starburst/AGN
connection and to the evolutionary scenario, triggered by interactions, proposed in our previous
papers. It is a follow-up spectroscopic pilot study aiming at investigating the possible
effects of interactions  on the neighbours of our Seyfert
galaxies and understanding the conditions necessary for the different types of
activity.

In \S 2 we will discuss our galaxy samples and we will present our observations
and data reduction. 
The spectroscopic analysis and classification of the galaxies, basic host galaxy properties, results from STARLIGHT stellar population synthesis code and the analysis
of the available X-ray 
observations are presented in section \S 3. Finally, in section \S 4 we
will interpret our results and draw our conclusions. All distances are calculated taking into account the local velocity field (which includes the effects of the following structures : Virgo, Great Attractor and Shapley) for the standard $\rm\Lambda$CDM cosmology ($\Omega_m$=0.27, $\Omega_\Lambda$=0.73). Throughout our paper we use $H_{\circ}=100 h$ km s$^{-1}$ Mpc$^{-1}$, following our previous study on the same samples.

\section{Data}
\subsection{Sample Definition and Previous Results}

The samples of active galaxies were initially compiled from the
catalogue of Lipovetskyj, Neisvestnyj \& Neisvetnaya (1987), which
itself is a compilation of all Seyfert galaxies known at the time from
various surveys and in various frequencies (optical, X-ray, radio,
infrared). It includes all extended objects and several starlike
objects with absolute magnitudes lower than -24. Available multifrequency data
are listed, including: coordinates,
redshifts, Seyfert type (and sub-type), UBVR-photoelectric
magnitudes, morphological types, fluxes in $\rm H\beta$ and [OIII]5007, JHKLN fluxes,
far-IR (IRAS) fluxes,
radio fluxes at 6 and 11 cm, monochromatic X-Ray fluxes in 0.3 - 3.5 and 2
- 10 keV. All data can be found online at the vizier database
(http://vizier.cfa.harvard.edu/viz-bin/VizieR?-source=VII/173). About
half of the listed Seyfert galaxies can also be found in the IRAS
catalogue.

Dultzin-Hacyan et al. (1999) selected from the catalogue two volume
limited and complete samples, consisting of 72 Sy1 and 60 Sy2, to
study their projected circumgalactic environment. In Koulouridis et al. (2006a)
we used practically the same samples in order to verify their results, using in
addition redshift data from the CFA2 and SSRS surveys and our own
deeper spectroscopic observations. 
%Our samples were limited in the area of these two surveys (48 Sy1 and
%56 Sy2) but were tested for consistency with the original ones. 
Well selected control samples
(same redshift, diameter and morphology distributions) were used for
the comparison in both studies.
%\begin{figure}
%\epsscale{0.8}
%\plotone{overal_bw.ps}
%\figcaption{Fraction of BIRGs (top panels, thick line), Sy1s (bottom panels,
%thick line), 
%Sy2s (thick dashed line) and their respective control 
%sample galaxies (thin lines) which have their neighbour 
%within the indicated redshift separation,
%as a function of projected distance. Uncertainties are 
%$\sim \pm 0.05$ in fraction (from Koulouridis et al. 2006b).}
%\end{figure}

Using the CfA2 and SSRS redshift catalogues, and our own
deeper low-resolution spectroscopic
observations (reaching to $m_B\sim 18.5$), we searched for neighbours 
within a projected distance $R\leq 100$ h$^{-1}$ kpc and a radial
velocity separation $\delta u\leq 600$km/sec and we found that:
\begin{itemize}
\item The Sy1 galaxies and their control sample show a similar (consistent within
1$\sigma$ Poisson uncertainty) fraction of objects having at least one close
neighbour.
\item There is a significantly higher fraction of Sy2 galaxies
having a near neighbour, especially within $D\leq 75$ h$^{-1}$ kpc,
with respect to both their control sample and the Sy1 galaxies.
%\item Results based on the BIRG sample, which includes mostly starburst and 
%Sy2 galaxies, show that an even higher fraction has a close neighbour.
\item The large-scale environment of Sy1 galaxies (D = 1 $h^{-1}$ Mpc
  and $\delta u\leq 1000$km/sec) is denser than that
  of Sy2 galaxies, although consistent with their respective control samples.
\item Using deeper spectroscopic observations of the neighbors for a random subsample
  of 22 Sy1 and 22 Sy2 galaxies we found that the differences between
  the close environment of Sy1 and Sy2's persists even when going to
  fainter neighbours, correspond to a magnitude similar to that of the
  Large Magellanic Cloud.
\end{itemize}

For the purposes of the present study we obtained new
medium-resolution spectroscopy, in order to resolve the $H\alpha$ and [NII]
lines - unresolved in our original low-resolution spectra, of
all the neighbours around the aforementioned subsamples
of the 22 Sy1 \& 22 Sy2, respectively. In Table 1 and 2
we present the names, celestial coordinates, $O_{MAPS}$ magnitudes
\footnote{$O$
(blue) POSS I plate magnitudes of the Minnesota Automated Plate
Scanner (MAPS) system. We use $O_{MAPS}$ magnitudes because Zwicky magnitudes were not
available for the fainter neighbours, and we needed a homogeneous
magnitude system for all our objects.} and redshifts of the 
Sy1 and Sy2 galaxies which have at least one close neighbour (within
$\delta u <600$km/sec). The full samples are presented in detail in Koulouridis et al. (2006).
Note that we have kept the original neighbours enumeration of
the previous papers (for example, in table 2, NGC1358 has only
neighbour 2, since neighbour 1 had $\delta u > 600$km/sec).

\subsection{Spectroscopic Observations}
We have obtained medium-resolution spectroscopic
data of all the neighbouring galaxies in our samples in order to classify
them according to their optical emission lines (\S 2.3).
Optical spectra were taken with the Boller \& Chivens spectrograph  
mounted on the  2.1m telescope at the Observatorio Astron\'omico  
Nacional in San Pedro M\'artir (OAN-SPM). Observations were carried out during  
photometric conditions. All spectra were obtained with a 2$\farcs$5 slit. 
The typical wavelength range  
was 4000-8000 \AA\ and the spectral resolution R=8\AA\ .   
Spectrophotometric standard stars were observed every night.

The data reduction was carried out with the IRAF\footnote{IRAF is  
distributed by National Optical Astronomy Observatories
operated by the Association of Universities for Research in  
Astronomy, Inc. under cooperative agreement with the National
Science Foundation.} package following a standard procedure. Spectra  
were bias-subtracted and corrected with dome
flat-field frames. Arc-lamp (CuHeNeAr) exposures were used for  
wavelength calibration. All spectra can be found in Appendix A.

\subsection{Analysis and Classification Method}
In this section we present results of our spectroscopic observations 
of all the neighbours with $D\le100\;h^{-1}$ kpc and $m_{O_{MAPS}}\mincir
18.5$ for the samples of Sy1 and Sy2 galaxies. We have also
used SDSS spectra when available. 

Our aim was to measure six emission lines: H$\beta \; \lambda
4861$, H$\alpha \; \lambda 6563$, [NII] $\lambda 6583$, [OIII] 
$\lambda 5007$, [SII] $\lambda 6716$ and [SII] $\lambda 6731$,
in order to
classify our galaxies, using the Baldwin, Phillips \& Terlevich (1981,
hereafter BPT) and Veilleux \& Osterbrock (1987) diagrams.
For the cases that it was not possible to measure the H$\beta$ and [OIII] emission lines,
we use the more approximate classification by Stasi{\'n}ska et al. (2006).

Based on the above, we adopted the following classification scheme:
\begin{itemize}
\item  absorption line galaxies (ALG), i.e. galaxies with no emission lines. 
\item galaxies with emission lines (ELG), meaning that they exhibit 
nuclear or/and recent star forming activity.
\end{itemize} 

Flux ratios for the emission lines mentioned above have been measured
after subtracting the host galaxy contamination from each
spectrum. We disentangle the spectral contribution of the host galaxy
from the observed spectra by  using the stellar population synthesis
code  \texttt{STARLIGHT}\footnote{http://starlight.ufsc.br/}.
Spectra processing and fits were carried in the same fashion as
described in section 3.1 of   Le\'on-Tavares et al. (2011). For a
detailed description of the \texttt{STARLIGHT } code  and its
scientific results,  we refer to the papers of the SEAGal
collaboration (Cid-Fernandes et al. 2005, Mateus et al. 2006; Asari et
al. 2007; Cid-Fernandes et al. 2007). We only note that we have calculated the 1$\sigma$
standard deviation of the flux as follows (Tresse et al. 1999):
\[\sigma=\sigma_cd\sqrt{2N_{pix}+EW/d}\,\,\,\,\,(1)\]
where $\sigma_c$ is the standard deviation of the continuum about the
emission line, $d$ is the spectral dispersion in \AA\ per pixel and
$N_{pix}$ is the base-width of the emission line in pixels. In our case the parameter $d \sim 4$\AA/pix, while for the SDSS spectra is $\sim 1.1$\AA/pix for the H$\beta$ area and $\sim 1.5$\AA/pix for the H$\alpha$ area.
To the above we have added in quadrature the errors of the Gaussian fitting of the emission lines. We should note here that in some cases the B telluric band is very close to the [SII] doublet (see for example NGC 1019-N2 on the left of the doublet or UGC 7064-N1B on the right of the doublet) introducing a further uncertainty on the calculation of the flux. In all such cases we have simultaneously fitted the telluric absorption and the emission lines in order to have a better measure of the [SII] doublet's flux. Although we do not have an exact evaluation of the uncertainty due to the above spectral feature, we presume (at least for the cases that the B telluric band is close to the doublet) that the reported error is underestimated. \footnote{We should also note that the standard deviation of the continuum about the [SII] doublet was calculated after the subtraction of the B telluric band.} 

Although it is possible to
distinguish between a Star Forming Nucleus (SFN) galaxy\footnote{We choose to call SFN all
galaxies with prominent emission lines that do not show AGN activity.} and an AGN using only the
[NII]/H$\alpha$ ratio, we cannot distinguish between a low ionization
(LINER) and a high ionization (Seyfert) AGN galaxy. We have also
measured [OI] ($\lambda=6300$) when possible, as an extra
indicator of AGN activity. However, the weakness of the line in most cases did
not allow further use of it in a separate BPT diagram. 
\begin{figure}
\centering
\resizebox{8cm}{8cm}{\includegraphics{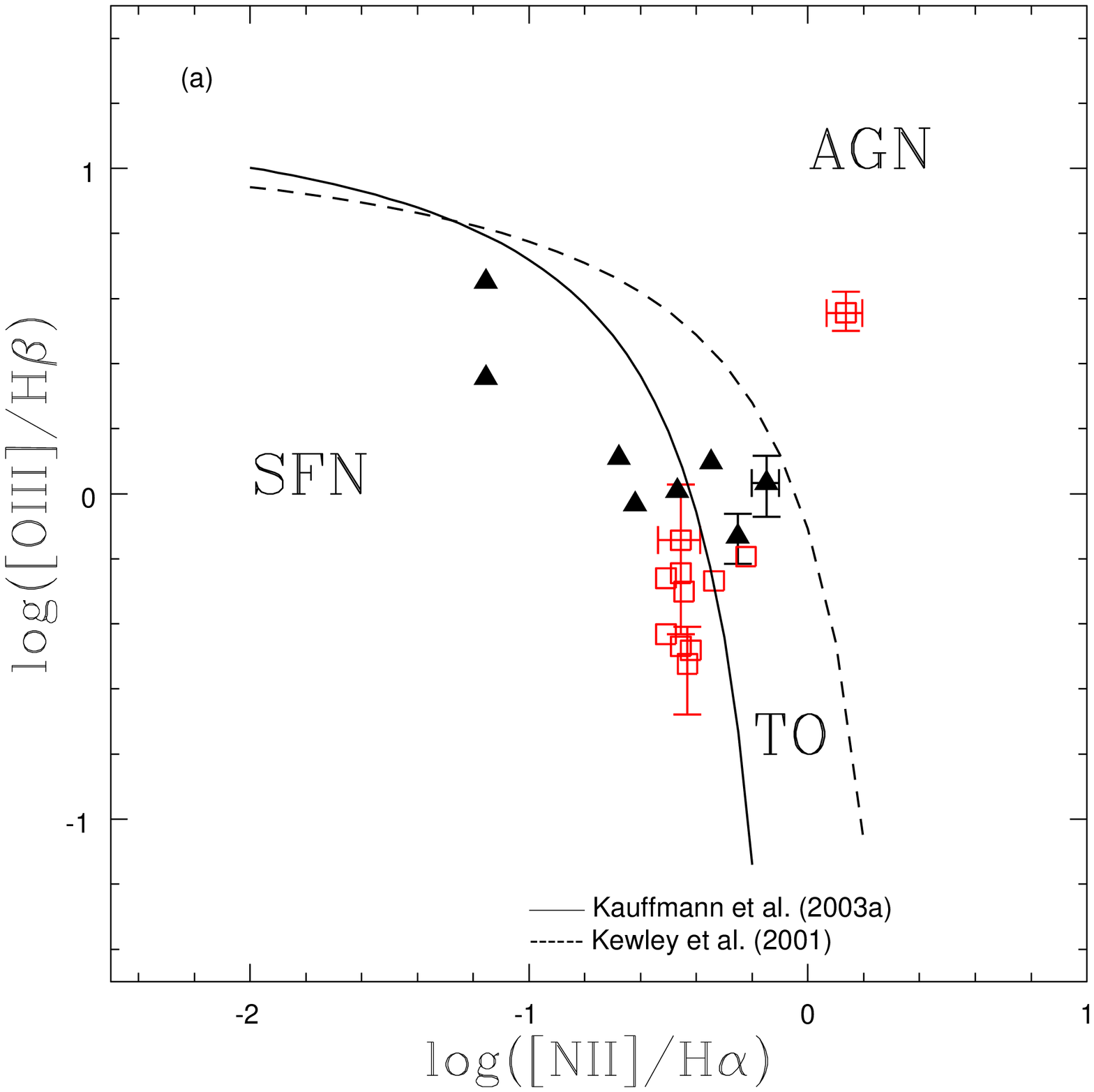}}
\resizebox{8cm}{8cm}{\includegraphics{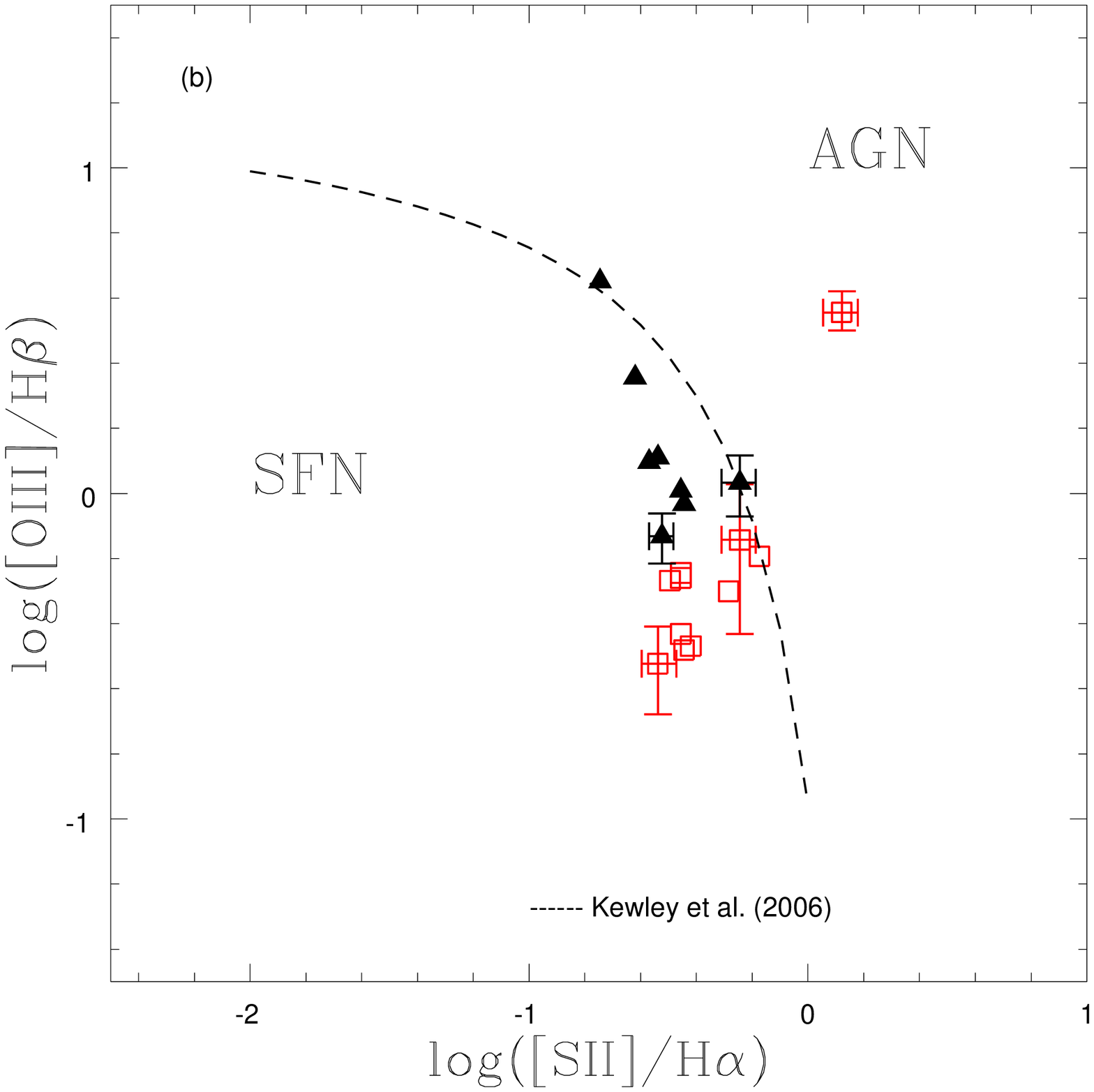}}
\caption{
(a) BPT diagram, and (b) Veilleux \& Osterbrock classification diagram of
the neighbours of Sy1 and Sy2 galaxies. The neighbours of Sy2 and Sy1 galaxies are
indicated by black triangles and red squares, respectively. For clarity, only the errorbars for those
galaxies with the largest uncertainties in the [OIII]/H$\beta$ ratio are presented in the diagrams.}
\end{figure}

In Fig.1a we plot the line ratios log([OIII]/H$\beta$) versus log([NII]/H$\alpha$)
(BPT diagram) for those neighbours of Seyfert galaxies for which we
have available the four necessary emission lines
%which we have obtained the full spectrum
\footnote{We have excluded one merger neighbour 
(UGC7064-N1) since its two nuclei are in an advanced merging state and
their properties are most probably independent of any interaction
which may have with the central active galaxy}. We also 
plot the Kauffmann et al. (2003a) separation line between SFN and AGN
galaxies, given by:
%$$\log([{\rm OIII}]/H\beta)=0.61/(\log([{\rm  NII}]/H\alpha)-0.05)+1.3\;,$$
$$\log([{\rm OIII}]/\rm H\beta)=\frac{0.61}{(\log([{\rm
NII}]/H\alpha)-0.05}+1.3\;,$$
and the corresponding one of Kewley et al. (2001): 
%$$\log([{\rm OIII}]/H\beta)=0.61/(\log([{\rm  NII}]/H\alpha)-0.47)+1.19 \;.$$
$$\log([{\rm OIII}]/\rm H\beta)=\frac{0.61}{(\log([{\rm NII}]/\rm H\alpha)-0.47}+1.19
\;.$$

We also plot in Fig.1b the line ratios log([OIII]$/\rm H\beta$) vs
log([SII]$/\rm H\alpha$). Qualitatively, the same results as those
presented in Fig.1a are repeated
here as well. The dividing line is given by Kewley at
al. (2006). However, we do not have the respective line of Kauffmann
et al. (2003a), as
it  is not available in the literature, and thus we cannot separate
pure star forming galaxies from
composite objects. Since, as we have already discussed, the measurement of the [SII] doublet is probably contaminated by absorption of the B telluric band, we will draw our results based on the [NII] forbidden line.

We can now classify our objects in the following categories: 
\begin{itemize}
\item SFN: all the objects which are found below the line of Kaufmann et al.
\item AGN: the objects which are found above the line of Kewley et al. 
\item TO (transition object): the ones that are found
between the two lines and exhibit characteristics of both nuclear activity and recent star formation.
\end{itemize}
We do not attempt to divide the star forming galaxies into more 
subcategories since such a categorization appears to be highly subjective 
and depends on the applied methodology (e.g. Knapen \& James, 2009).

Note that for one of our objects (ESO 545-G013-N1) the $\rm H\beta$ and [OIII] $(\lambda5007)$ lines were not observed and therefore we classified it using the more approximate method of Stasi{\'n}ska et al. (2006), which is based solely on the NII/H$\alpha$ ratio.
in order to evaluate this method, we applied it to all our galaxies and found a consistency with the BPT classification in all cases but one (see Table 3).

Further classification of the Seyfert galaxies in type 1 and type 2 was
obtained by direct visual examination of the spectra from the broadening of the
emission lines. No broad lines were discovered in the spectrum of the two neighbours classified as AGN 
and therefore they should be considered as type 2.
In Tables 1 and 2 we list, for all neighbours, their line ratios and the two different
classifications.

We also measured the equivalent width of the H$\alpha$ emission
line, in order to use it as an extra indicator of the galaxy's star forming history, in addition with the STARLIGHT code's results. The minimum Equivalent Width, defined as the integrated local continuum rms noise normalized to the level of the local continuum, at a 5$\sigma$ confidence level, is found to be EW$_{min}\sim 1$\AA.
%and all detected lines have at least $3\times\rm EW_{min}$
We should note here that in a small number of cases the [OIII] and H$\beta$
lines were detected only after the subtraction of the continuum. We
calculated the 1$\sigma$ standard deviation of the EW as follows
(Tresse et al. 1999):
\[\sigma_{EW}=\frac{EW}{F}\sigma_cd\sqrt{2N_{pix}+EW/d+(EW/d)^2/N_{pix}}\,\,\,\,\,(2)\]
where $\sigma_c$ is the standard deviation of the continuum about the
emission line, $d$ is the spectral dispersion in \AA\ per pixel,
$N_{pix}$ is the base-width of the emission line in pixels and F the
flux of the emission line. 

\section{Results and analysis.}
\subsection{Activity of the neighbours.}
In this section we discuss in more detail the results of our 
spectroscopy and classification. We have excluded the merging neighbour of UGC 7064, since 
the properties of its two nuclei are more affected by their mutual
interaction rather than 
by their neighbouring Seyfert.
We can draw our first results for 
each sample separately inspecting Table 1 and 2.
From the analyzed 15 
neighbours of Sy1 only 4 are ALGs, while 8 of them are
SFNs, 2 are classified as TOs and one is classified as AGN. Similar results
hold for the neighbours of Sy2 galaxies. 4 out of 13 neighbours do not
present emission lines, 6 are SFNs and 3 are TOs.
Therefore, at least 70\% of the neighbours, within $100 \; h^{-1}$ kpc, of both type of
Seyfert galaxies have emission lines. We should note
here that Ho et al. (1997), studying a magnitude limited sample of
galaxies ($B_T\leq$12.5), came up with a similar high percentage of
activity (86\%). However, the results of our sample of faint
neighbours cannot be directly compared with those of Ho et al. due to
the brighter magnitude limit of the latter.

We can extract one of the most interesting results of our analysis by
examining Fig.1, i.e., that the neighbours of Sy2 galaxies have
systematically larger values
of [OIII]$/\rm H\beta$ than the neighbours of Sy1 galaxies. Using a
Kolmogorov-Smirnov two-sample test for the [OIII]$/\rm H\beta$ ratio
we find that the null hypothesis that the samples are drawn from the
same parent population is rejected at a 99.9\% level. Especially for
those galaxies that exhibit only star-formation, the ratio
[OIII]/H$\alpha$ is mainly related to their ionization level. This
fact could be an indication of a more recent starburst event in the
neighbours of Sy2 galaxies than of Sy1's, caused possibly by the
interaction with a neighbouring galaxy, or an effect of the galaxy
downsizing i.e. more massive galaxies have formed their stellar
populations earlier than less massive ones (Asari et al. (2007) argue that the location of galaxies on the BPT diagram is considered to be a result of downsizing).
%On the other hand, as we can derive from Table 3, the SFN neighbours
%of the Sy2 galaxies tend to have relatively high values of absolute
%magnitude $M$ (four out of six having the highest), probably
%indicating lower stellar masses.
Should the downsizing explanation be true, the ionization level can be
considered as an indicator of metallicity, which is closely related to
the stellar mass. Thus, galaxies having lower values of
[OIII]/H$\beta$ would be more massive and would have higher
metallicities, indicative of an older average age of the stellar
population. In Table 1 and 2 we can see a weak trend of the mean
stellar metallicity ($<Z>$) values (extracted from STARLIGHT) for the
Sy2 SFN neighbours being lower with respect to that of the Sy1's and although
stellar masses cannot be directly derived from our data, most low
metallicity SFNs are also faint and small in size (Table 3). However,
no trend can be found by comparing the average age of the stellar
populations, and given the small number of galaxies these results remain rather
inconclusive. 
%The average age of the stellar population also changes along the
%[OIII]$/\rm H\beta$ axis, the higher ionized galaxies having higher
%ratio of current to past star-formation rates (Asari et al. 2007, Cid
%Fernandes et al. 2009). 
%Summarizing all of the above we conclude that the neighbours of
%Seyfert 2 galaxies show higher ionization, lower metallicity, less
%stellar mass and more importantly younger stellar populations than
%those of Seyfert 1 galaxies. Using a Kolmogorov-Smirnov two-sample
%test for the [OIII]$/\rm H\beta$ ratio we have verified that the null
%hypothesis, that the samples are drawn from the same parent
%population, is rejected at a 99.9\% level.

%The average age of the stellar population also changes along the
%[OIII]$/\rm H\beta$ axis, the higher ionized galaxies having higher
%ratio of current to past star-formation rates (Asari et al. 2007, Cid
%Fernandes et al. 2009). 

%In order to explore the possibility that the above trends are due to
%interactions of the central AGN with the companion galaxies, we study
%the circumnuclear star formation (a.) by measuring the ThEquivalentnt
%Width of the $\rm H\alpha$ emission line (EW($\rm H\alpha$))and (b.)
%by inspecting the data of the stellar population synthesis code.

The Equivalent Width of the $\rm H\alpha$ emission line is also a good
indicator of the star formation history, since it represents the ratio
of present to past star formation, i.e. during a starburst event young
massive stars strengthen the emission lines and enhance their EW, but
as time passes the strength of the emission line fades, the continuum
rises again and the value of the EW declines. The highest values of
the EW($\rm H\alpha$) can be found in the spectra of the Star Forming 
neighbours of our Sy2 sample, while on the other hand some of the lowest values
can be found in the respective Sy1's neighbours spectra. 
%Adding the TO neighbours to the previous analysis (presuming the AGN
%contribution is small) the trend is enhanced (Tables 1 \& 2).

A more direct way to to explore the possibility that the differences
of the ionization level is due to the age of the interaction of the
central active galaxies with its neighbour, is by the determination of the age of the
most recent peak of star formation with the "STARLIGHT" code. As it
was expected however, most of the star forming galaxies present a
recent event within the last 20 Myr, a necessary fact in order to
detect strong emission lines and we can not detect any significant
differences between Sy1 and Sy2 SFN neighbours. On the other hand, an
interesting result is the fact that six out of seven Sy2's non-SFN
(ALG, AGN or TO) companions present a recent star formation peak $<$30
Myr, while six out seven Sy1's corresponding neighbours are "quiet"
for more than 100 Myr. The above fact may indicate that indeed the Sy1
galaxies have interacted with their neighbour earlier than the Sy2s.

Summarizing our main results of this section:
\begin{itemize}
\item More than 70\% of the neighbours of the two AGN samples exhibit optical
  emission lines, indicating recent star formation and/or nuclear activity.
\item Around 30\% of the neighbors of Sy1 and Sy2 galaxies show the presence
of AGN activity, mainly in the form of TOs.
\item The neighbours of Sy2s are systematically more ionized than the
neighbours of Sy1s and their EW(H$\alpha$) values tend also to be higher.
\item Most of the non-SFN neighbours of Sy2 galaxies show a recent
  starburst event ($<$30 Myr), while the corresponding age for most of
  the Sy1's neighbours is $>$100 Myr.
\item the previous two results indicate differences in the star
  formation history of the neighbours of different types of AGN as
  well as in the age of the most recent interaction.
%\item No broad lines (type 1 activity) were detected in the two neighbours which exhibit pure AGN spectra.
\end{itemize}

%Our results show a statistically significant difference of the
%ionization level between the neighbours of our Sy1 and Sy2
%samples. Some interesting trends were also discovered in our attempt
%to explain this result, but given the volume of our samples the
%interpretation remains elusive. We will discuss further this issues
%in \S 4.

Finally we should note how close to a
composite state are the neighbours of active galaxies, in agreement
with Kewley at al. (2006a) who showed
that the star forming members of close pairs, lie closer to the
classification line than the star forming field galaxies.
We suggest that galaxies between the curves of Kauffmann
et al. (2003) and Kewley et al. (2001) possibly migrate from a pure
star forming phase to
a pure AGN phase. This suggestion is of great importance to the
formulation of a possible 
evolutionary scenario and will also be discussed further in \S4.

\subsection{Magnitude and distance analysis}

Since we have already applied a homogeneous magnitude system to our samples,
we can now study whether there is a correlation between the activity
of an interacting pair of galaxies and their magnitudes.
The activity-magnitude comparison is performed by examining the
absolute magnitude difference between the neighbour 
and the central active galaxy ($\Delta M$), 
%since it could be considered a good tracer of the strength of the interaction, 
with small values ($\Delta M <1.5$) indicating a stronger pair interactions (we tag these
pairs as equally bright).
A further parameter that can be used is the absolute magnitude of
the neighbour, indicating its size. On average, absolute 
magnitude and size are correlated in small redshift intervals (as it is in our
case) and therefore we can safely presume that a faint galaxy 
is also small in size and a bright one is large. The latter has been also optically
inspected for our galaxies to further confirm the correlation (see
also maps of Fig.3), while the median absolute magnitude $M=-17.49$ is
considered to be the separating limit between bright and faint
companions. In addition we also examine the isophotal diameters at
25.0 B-mag arcsec$^{-2}$) from the Third Reference Catalogue of bright
galaxies (RC3) to compare with the absolute magnitudes, by considering
any neighbour with $D/D_{AGN}<1/2$ as being small. In two cases,
because of lack of RC3 data, their near-infrared isophotal diameters
(at 20.0 K-mag arcsec$^{-2}$) from the "Two Micron All Sky Survey Team
2MASS Extended Objects" (2MASS) catalogue) were used for the
comparison. We should note that only in the case of NGC 1241 the
diameter criterion is not in agreement with the absolute magnitude
criterion (marginally) and by inspecting also the SDSS image we
concluded that the neighbour is indeed small.
Finally, radial separation can also be considered as a crucial factor
of the strength of the interaction.
In Table 3 we list all the above mentioned values plus three
respective indices than take values between 0 and 1. With 1 we denote a
value that is in favor of the interaction, with 0 the opposite. In
more detail, if the radial separation $R$ is less than 50$h^{-1}$kpc
the respective index $I_D$ is 1 and the same is true for bright
neighbours and equally bright pairs, since all these factors may
affect positively the interactions between two galaxies. The sum of
the three indices is also listed in Table 3. Obviously the strength of
the interaction of a neighbour with the sum of the three indices equal
to 0 (i.e. small and faraway
neighbour of a large AGN) would be significantly different from one
with a sum equal to 3 (i.e. large and close galaxy of a comparable sized AGN).
It therefore becomes evident that:

\begin{itemize}
\item All faint neighbours and all neighbours of a non-equally
bright pair of galaxies are preferentially absorption line or purely
SFN. %star forming galaxies.
\item All neighbours which host an AGN or are transition objects (TO), fall in
the bright category and are neighbours of an equally bright pair.
\item All neighbours with interaction indices sim$\leq 1$ are purely star forming galaxies.
\item All ALGs, AGN and TO galaxies have interaction indices sum$\geq 2$ (except NGC863-N1).
\end{itemize}

From our results we can infer that when a faint/small galaxy comes in interaction with another galaxy, 
the encounter induces at most a starburst but no AGN activity in the small galaxy; however it can trigger a bright AGN in the larger one. This could be due to the absence in small galaxies
of a massive black hole (Wang, Kauffmann 2007; Volonteri et al. 2008). 
If this assertion is correct, only galaxies which experience a major close
interaction or merger can exhibit AGN
activity and this could be the reason why AGN hosts are more frequently found in
early type galaxies (e.g. Marquez \& Moles 1994; Moles, Marquez, \&
Perez 1995; Ho et al. 1997; Knapen et al. 2000; Wake et
al. 2004). This can also account for the large fraction of star forming
galaxies among our samples of neighbours.

To cover all aspects of this issue, we should mention here that Galaz
et al. (2011) showed that the fraction of low surface brightness 
galaxies hosting an AGN is significantly lower than the corresponding fraction
of high surface brightness galaxies, independently of the mass. 
So the deficiency of AGN in faint galaxies seems to be due to an intrinsic
inability of these galaxies to host or to feed a massive black hole.

Our results indicate that the interaction of a bright galaxy
especially in an equally bright pair results in an AGN or an ALG.
Finding some massive galaxies, members of an equally bright
interacting pair, without emission lines implies either a non-eventful
interaction or a delay of the outcome of the interaction. On the other hand, weak star formation 
or low luminosity nuclear activity may not be detectable by optical
spectroscopy, although it could possibly be detected in X-rays. Such
an analysis is presented below.
%In the X-ray analysis section below we address again this issue.

%We note that interactions may be an important
%ingredient in triggering activity, but certainly are not a sufficient
%condition. The morphology of the host galaxy, the mass of the central black hole
%or even more complicated issues, like the spin of the BH or the
%strength of the AGN feedback, could also play a role to the viability
%of the starburst phase and the feeding of the black hole. 

\subsection{The XMM-Newton observations} 
We explore here, using the {\it XMM-Newton} public
archive, whether the neighbours show X-ray
activity. We find that 13 target fields have been 
observed by {\it XMM-Newton}. However,  some of them are very bright 
 and have been observed in partial window mode, rendering the 
 observations in center of the Field-of-View unusable (NGC5548, NGC863,
1H1142-178, NGC7469). 
 The list of the remaining observations (13 neighbours and 9 central
 Seyfert galaxies) is shown in Table 4, in which
we present X-ray fluxes for the detections  as well as upper limits for
the non-detected sources. The fluxes have been taken from 
 the 2XMM catalogue  (Watson et al. 2009).
 The fluxes refer to the total 0.2-12 keV band  
 for the PN detector or the combined MOS detectors in the case 
 where PN fluxes are not available and are estimated using a 
 photon index of $\Gamma=1.7$ and an average Galactic column density of 
 $\rm N_H = 3\times 10^{20}$ $\rm cm^{-2}$. Luminosities were
 estimated using the same spectral parameters.
  In the same table we quote the 2XMM hardness ratios,
  derived from the  1-2 keV and 2-4.5 keV bands (hardness ratio-3
  according  to the 2XMM catalogue notation). 
 The upper limits, derived using the {\sl FLIX} software, are
 estimated following the method of Carrera et al. (2007). 
 This provides upper-limits to the X-ray flux at a given 
 point in the sky covered by {\it XMM-Newton} pointings.  
The radius used for deriving the upper limit was 20 or 30 arcsec
 depending on the presence of contaminating nearby sources.

In Fig.2 we present the X-ray to optical flux diagram
$f_X-f_B$  (e.g. Stocke et al. 1991).
\begin{figure}
\resizebox{8cm}{8cm}{\includegraphics{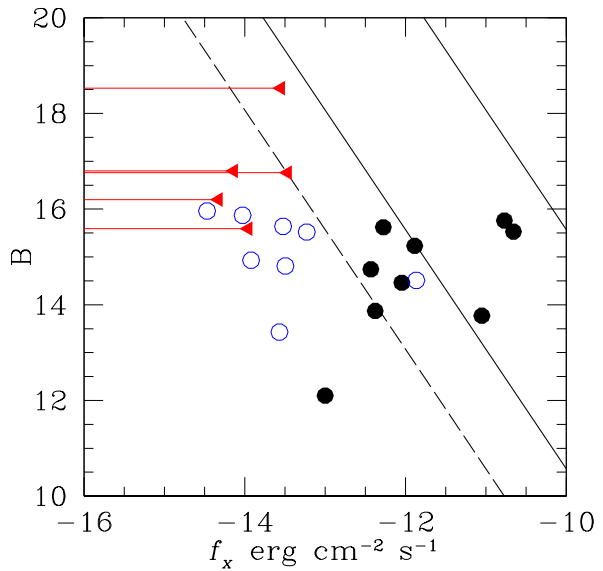}}
    \caption{The X-ray (0.2-12 keV) to  optical (B-band) flux
      diagram for both the central active galaxy targets (solid circles) and the neighbors 
     (open circles). The triangles (upper limits) denote the neighbors
     with no X-ray detection. The upper, lower solid line and the dash
     line correspond to $f_X/f_B=+1, -1, -2 $ respectively. The only neighbour (open circle) that lies in the AGN regime is NGC 7682-N1}
    \end{figure}
This diagram provides an idea on whether a galaxy may host an active nucleus. 
This is because AGN have enhanced X-ray emission for a given optical magnitude 
relative to ALG galaxies. The space usually populated by AGN is shown
between the continuous lines. The central Seyfert galaxies are shown
as filled points, but since X-ray flux has not been corrected for
X-ray absorption,
a number of absorbed AGN galaxies lie between the lower continuous line and the dashed line, while the
heavily absorbed Sy2 NGC 7743 (Akylas \& Georgantopoulos 2009), lie far below the dashed line.
One neighbour which lies in the AGN regime (NGC 7682-N1) can be clearly
seen. This has been classified as a TO galaxy in the
optical spectroscopic analysis and is one of the three
neighbours (for which XMM-Newton observations are available, see Table 4) 
having an active nucleus based on optical spectroscopy.
Additional information on the nature of our sources can be extracted from the
hardness ratios. 
Two sources NGC 526-N2 and NGC 1358-N2 have hardness ratios suggesting an
absorption of $\rm N_H \approx 10^{22}$ $\rm cm^{-2}$, 
consistent with the presence of a moderately obscured active
nucleus. Both these galaxies present no optical emission lines and thus are
classified as ALG, based on their optical spectra. In other words, the
lack of optical emission lines from the
nucleus of these objects could be a result of obscuration and indeed this seems to be the case, since the detection limit of the EW of emission lines is low enough. In
addition, we should mention here that all galaxies among those which fall in fields 
observed by XMM-Newton, classified as ALG through
optical spectroscopy, present X-ray emission. 
%possibly also indicating 
%the presence of nuclear activity. 
In contrast, all SF galaxies except one
in the X-ray subsample do not show an X-ray detection. 

We should note here that unobscured low accretion rate Sy2 objects
and/or low luminosity AGN, where the Narrow Line Region (NLR) cannot
be detected by means of optical spectroscopy, or even X-ray binaries
may account for the X-ray detection of unobscured ALG
galaxies. However, emission from X-ray binaries is not detected in the
spectra of the SFNs rendering this interpretation less plausible. This
analysis therefore implies that the total fraction of neighbours of
AGN that show recent star formation or AGN, based on optical
spectroscopy or X-ray observations, is at least $80\%$ and possibly
quite higher. This matter will be fully addressed in future work.

\section{Discussion \& Conclusions}
In this paper we investigate the close environment ($\leq
100 \; h^{-1}$kpc) of a local sample ($z<0.034$) of
AGN. In particular we explore the spectroscopic, photometric and X-ray properties of 30 neighbouring galaxies
around 10 Sy1 and 13 Sy2 galaxies.
Based on optical spectroscopy, in our current study we have found that
the large majority of these neighbours show some activity, mostly recent
star-formation (emission line spectrum) but AGN as well. In addition,
our X-ray analysis of a subsample of neighbours with public XMM-Newton
observations showed that the neighbours which are classified as ALG
based on optical spectroscopy might have a low luminosity active core,
since all of them are X-ray detected, while two out of five appear to
have a moderately obscured active nucleus. The X-ray detections could
be due to X-ray binaries, but we argue that this is less probable
since the pure star forming neighbours do not show any X-ray emission
down to the flux limit of the available observations. From both
optical spectroscopy and X-ray observations, it becomes clear that the
fraction of AGN's neighbours which exhibit recent star formation
and/or nuclear activity, within 100 $h^{-1}$ Mpc, is $>80\%$ and
possibly higher.

Furthermore, the close neighbours of Sy1 galaxies, especially the
SFNs, are less ionized and have lower values of EW(H$\alpha$) with
respect to those of Sy2 and thus seem to be a different, more evolved
population than those of Sy2s. Other discovered trends of metallicity,
host galaxy size and age of the most recent starburst event indicate
possible physical differences between the neighbours of Sy1 \& Sy2
galaxies as well, a fact that may link AGN activity with interactions.

Indeed, over the past two decades there have been several studies
which supported that the idea of an evolutionary
sequence from starburst to Seyfert galaxies (e.g. Storchi-Bergmann et
al. 2001, see also introduction). Furthermore, there are also studies
that separate type I from type II objects (e.g. Hunt et al. 1997;
Maiolino et al. 1997, Gu et al. 2001) implying that recent
star-formation is only present in type II objects (see also Coldwell
et al. 2009). Based on the number and proximity of close ($\mincir
60-100\;h^{-1}$kpc) neighbours, around different types of active (Sy1,
Sy2 and BIRG) galaxies (e.g. Dultzin-Hacyan et al. 1999; Krongold et
al. 2002; Koulouridis et al. 2006a,b), a very interesting evolutionary
sequence has been suggested, starting with a close interaction that
triggers the formation of a nuclear starburst, subsequently evolving
to a type 2 Seyfert, and finally to a Sy1. Recent observational results by Villarroel et al.(2012) and Kollatschny et 
al.(2012) also seem to support this scheme.
This sequence is likely independent of luminosity, as similar trends have been proposed for
LINERs (Krongold et al. 2003) and ULIRGs and Quasars (Fiore et
al. 2008 and references therein). The above findings were also supported by numerical simulations (Hopkins et al. 2008) which outlined such an evolutionary scheme for
merging galaxies. The proposed activity evolution can explain
the excess of starbursts and type 2 AGN in interacting systems, as
well as the lack of type 1 AGN in compact groups of galaxies
(Mart{\'{\i}}nez et al. 2008) and galaxy pairs (e.g. Gonzalez et
al. 2008).

Since the physical properties of the neighbours should be reflected on
the state of the central active galaxy,
we argue that our results may be in the same direction as those of our
previous papers (Koulouridis et al 2006a,b), supporting an
evolutionary sequence of galaxy activity, driven by interactions, the
main path of which follows the sequence of induced star formation, Sy2
and finally Sy1 phase. A time delay should exist between the pure star
forming and AGN phase (see discussion in the introduction), where
active nucleus and circumnuclear starburst coexist. In this initial
phase, the nucleus is heavily obscured by the still star forming
molecular clouds and it can be observed as a transition stage of
composite Sy2-starburst objects. We should note here that according to
Ballantyne, Everett \& Murray (2006) a non-evolving torus cannot
provide the AGN obscuration over all cosmic time and that extra
obscuration by star formation is needed.

The most probable manner for the AGN to dominate is to eliminate the
starburst, possibly by the AGN outflows or by radiation pressure. We
point out that a great theoretical success of the starburst/AGN
connection is the quenching of the induced star formation by the AGN
feedback, which can explain the formation of red and dead elliptical
galaxies (e.g. Springel et al. 2005a; Di Matteo et al. 2005; Khalatyan
et al. 2008). This can be achieved by outflows from the core which
have enough energy to dissipate the material around it and thus
suffocate star formation (e.g. Krongold et al. 2007, 2009; Blustin et
al. 2008; Hopkins \& Elvis 2010; Novak, Ostriker \& Ciotti 2011;
Cano-D{\'{\i}}az et al. 2012; Zubovas \& King 2012). Recent
observational studies and simulations have shown that AGN's ionized
outflows may carry enough energy to cease star formation in the host
galaxy rapidly, in less than 1 Gyr (see for example Kaviraj et al. 2011). As
the starburst fades (see relevant discussion and references in the introduction), the Seyfert 2 
state starts dominating, to be followed at the end by a totally
unobscured Sy1 state, plausibly $\sim 1$ Gyr after the initial
interaction (see Krongold et al. 2002). More details about the
co-evolution of the torus and the AGN are given by Liu \& Zhang (2011),
supporting our evolutionary scheme. We should note here that recent
observations (Hasinger et al. 2008; Treister et al. 2010) verified a
significant increase of the type 2 AGN fraction with redshift, a fact
which is in agreement with our evolutionary scheme.

%We now attempt to interpret our present results within the above evolutionary scheme. 

% The time needed for type 1 activity to appear should be larger than the
%timescale for an
%unbound companion to escape from the close environment, or comparable
%to the timescale needed for an evolved merger ($\sim 1$Gyr, see
%Krongold et al. 2002).

Alternatively, there is a possibility that the SFN neighbours of Sy1
galaxies are systematically more 
massive with respect to those of Sy2 and that their older stellar
population is due to downsizing, i.e. more massive galaxies have
evolved earlier, while less massive ones exhibit more recent star
formation and thus younger stellar population. However, there is no
obvious explanation on why more massive galaxies should be located
preferentially near Sy1 galaxies and not Sy2. The combination of both
downsizing and the interaction driven sequence, as presented
previously, can also be at work.

We stress that the suggested evolutionary scenario does not
invalidate completely the unification scheme. It implies that the orientation of the torus can
determine the AGN phenomenology only at specific phases of the evolutionary sequence. 
In particular, this probably occurs when the obscuring
molecular clouds form the torus (possibly when the AGN activity
reaches its peak $\sim$0.5 Gyr after the initial interactions (Kaviraj
et al. 2011)) and before being completely swept away (possibly
after 1 Gyr (Krongold et al. 2002)).
From our point of view, in an ever evolving universe an evolutionary
scheme, is more probable than the original
unification paradigm which proposes a rather static view of AGN. Of
course, orientation could and
should also play a role between the obscured Sy2 and Sy1 phase, when
the relaxing obscuring material forms a toroidal structure. 

There are still many unresolved issues and caveats concerning 
these suggestions, since the evolutionary sequence is not unique and should also
depend  on the geometry, the density and other factors of the obscuring and 
the accreting material, as well as on the mass of the host galaxy and its 
black hole. Furthermore, the sample presented in this pilot study is
rather small and the results should be considered as indicative and
should be confirmed by analysis of larger samples.

\subsection*{Acknowledgments}
EK thanks the IUNAM and INAOE, were a major part of this work was
completed, for their warm hospitality. We also thank OAGH and OAN-SPM staff for excellent assistance and  
technical support at the telescopes. 
VC acknowledges funding by CONACyT research grants 54480
and 15149 (M\'exico).
MP acknowledges funding by the Mexican Government research grant
No. CONACyT 49878-F and DD
support from grant PAPIIT IN111610 from DGAPA, UNAM.
YK acknowledges support from CONACyT 168519 grant  and UNAM-DGAPA PAPIIT IN103712 grant.
This research has made use of the USNO-B catalog (Monet et al. 2003)
and the MAPS Catalog of POSS I (Cabanela et al. 2003) supported by the
University of Minnesota (the APS databases can be accessed at
http://aps.umn.edu/).
The STARLIGHT project is supported by the Brazilian agencies CNPq,
CAPES and FAPESP and by the FranceBrazil CAPES/Cofecub programme.
Funding for the SDSS and SDSS-II has been provided by the Alfred P. Sloan
Foundation, the Participating Institutions, the National Science Foundation,
the U.S. Department of Energy, the National Aeronautics and Space
Administration, the Japanese Monbukagakusho, the Max Planck Society, and the
Higher Education Funding Council for England. The SDSS Web Site is
http://www.sdss.org/.
Finally, we would like to thank the anonymous referee for his or hers comments and suggestions that helped to significantly improve our paper.

\begin{landscape}
\begin{table}
\centering
\begin{minipage}{220mm}
\caption{Observational \& SSP results, emission line ratios and classification.}
\tabcolsep 3pt
\begin{tabular}{lccccccccccccccccc}
{\em NAME} &{No} & {\em RA}       &{\em DEC}     & m  &{\em U.T.} & {\em start U.T.} &  {\em exp.}      & {\em $\rm [OIII]/\rm H\beta$}      
&{\em $\rm [NII]/\rm H\alpha$} & {\em $\rm [SII]/\rm H\alpha^\star$}& EW(H$\alpha$)& $\chi^2$ & $<Z>$ &$\rm <logt>$&$\rm logt_{SB}$ 
                         &{$\rm C_{St}$} & { $\rm C_{\rm BPT}$}\\
{\em (1)}&{\em (2)}&{\em (3)}&{\em (4)}&{\em (5)}&{\em (6)}&{\em (7)}&{\em (8)}&{\em (9)}&{\em (10)}&{\em (11)}&{\em (12)}&{\em (13)}&{\em (14)}&{\em (15)}&{\em 16)}&{\em (17)}&{\em (18)}\\
\hline
&&&&&&&&\\
NGC 863                &   & 02 14 33.5 & $-$00 46 00 & 14.58   &&&&               &\\
                  &N1  & 02 14 29.3 & $-$00 46 05 & 18.25  &SDSS & - &-      & - & - & - & - & 0.6 & 0.020 &9.30&9.30 &ALG & ALG      \\
MRK 1400                 & & 02 20 13.7 &   +08 12 20 & 17.07    &&&&              &\\
                 &N1  & 02 19 59.8 &   +08 10 45 & 17.25 & 06/10/07 & 07:31 & 4800    & 0.55$\pm0.04$ & 0.31$\pm0.01$ & 0.35$\pm0.01$  &-37.4$\pm0.8$&1.3&0.012&9.07&6.70& SFN & SFN      \\
NGC 1019              &    & 02 38 27.4 &   +01 54 28 & 15.02                  &\\
                &N2  & 02 38 25.4 &   +01 58 07 & 16.28 & 21/10/06 & 08:25 & 4800   & 0.54$\pm0.09$ & 0.46$\pm0.01$ & 0.32$\pm0.01$ &-19.6$\pm0.3$&1.8&0.008&8.38&6.93& TO& TO        \\
NGC 1194            &      & 03 03 49.1 & $-$01 06 13 & 15.38\\
                  &N1  & 03 03 41.2 & $-$01 04 25 & 16.99 & SDSS & - & - & 0.37$\pm0.05$ & 0.31$\pm0.01$ & 0.35$\pm0.01$ &-20.9$\pm0.3$&0.8&0.039&8.16&6.30 &SFN & SFN      \\
                  &N4  & 03 04 12.5 & $-$01 11 34 & 15.75 &25/10/06 & 07:56 & 5400 & 0.33$\pm0.04$ & 0.38$\pm0.01$ & 0.36$\pm0.01$ &-20.4$\pm0.4$&1.3&0.017&8.64&7.20& SFN& SFN      \\
1H 1142$-$178     &        & 11 45 40.4 & $-$18 27 16 & 16.82                  &\\
       &N1   & 11 45 40.9 & $-$18 27 36 & 18.01 &19/05/07 & 04:29 & 3000 &-&-&-&-&1.4&0.014&9.38&8.40& ALG & ALG \\
       &N2   & 11 45 38.8 & $-$18 29 19 & 18.45 &21/05/07 & 04:16 & 6000 &0.72$\pm0.35$& 0.35$\pm0.06$ &0.57$\pm0.08$ &-8.0$\pm0.7$&2.4&0.050&9.11&9.10&SFN & SFN \\
MRK 699         &          & 16 23 45.8 &   +41 04 57 & 17.21                  &\\
         &N1  & 16 23 40.4 &   +41 06 16 & 17.59 &18/05/07 & 10:27 & 2100 &0.64$\pm0.24$& 0.60$\pm0.07$ & 0.67$\pm0.06$  &-8.2$\pm0.5$&2.3&0.030&9.31&9.11& TO & TO \\
NGC 7469              &    & 23 03 15.5 &   +08 52 26 & 14.48                  &\\
   &N1   & 23 03 18.0 &   +08 53 37 & 15.58 &01/12/06 & 03:07 & 3600& 0.30$\pm0.09$ & 0.37$\pm0.01$ & 0.29$\pm0.01$  
&-31.7$\pm0.5$&1.0&0.010&7.51&6.90&SFN & SFN \\
NGC 526A\tablefootmark{2}&& 01 23 54.5 & $-$35 03 56 &
15.69\tablefootmark{3} &\\
 &N1 & 01 23 57.1 & $-$35 04 09 &
15.80\tablefootmark{3}    &08/10/07 & 06:39 & 2400& 3.60$\pm0.50$ & 1.37$\pm0.20$ & 1.32$\pm0.19$ &-3.0$\pm0.7$&1.9&0.036&10.21&10.30&AGN&
AGN \\
 &N2  &01 23 58.1 & $-$35 06 54 &
15.68\tablefootmark{3}   &08/10/07 & 08:39 & 1500&-&-&-&- &1.4&0.030&9.93&8.07&ALG & ALG \\
 &N3 & 01 24 09.5 & $-$35 05 42 &
16.37\tablefootmark{3}    &08/10/07 & 09:34 & 3600& 0.57$\pm0.13$& 0.35$\pm0.02$ & 0.35$\pm0.03$ &-31.1$\pm0.6$ &1.9&0.019& 9.57&9.39&SFN
& SFN \\
 &N4  & 01 23 59.2 & $-$35 07 38 &
16.04\tablefootmark{3}  &08/10/07 & 07:33 & 3600& 0.34$\pm0.10$ & 0.34$\pm0.01$ & 0.38$\pm0.01$  &
-20.4$\pm0.5$&1.1&0.026&8.33&7.11&SFN & SFN \\
NGC 5548         &         & 14 17 59.5 &   +25 08 12 & 14.18                  &\\
        &N1 & 14 17 33.9 &   +25 06 52 & 17.16    &SDSS&-&-& 0.50$\pm0.20$ & 0.36$\pm0.02$ & 0.52$\pm0.02$  &
-9.4$\pm0.2$&0.5&0.026&8.00&6.39&SFN & SFN \\
NGC 6104          &        & 16 16 30.7 &   +35 42 29 & 15.11                  &\\
          &N1  & 16 16 49.9 &   +35 42 07 & 16.44   &18/05/07 & 09:37 & 1800 &-&-&-&-&2.3&0.050&9.11&9.10&ALG & ALG \\
&&&&&\\
\hline
\end{tabular}
\tablefoot{{\it(1)} name of AGN, {\it(2)} number of neighbour, {\it(3)-(4)} right ascension and declination in the equatorial coordinate system, {\it(5)} {\em $O_{\rm MAPS}$ apparent magnitude}, {\it(6)-(8)} date (dd/mm/yy), time and total exposure time (sec) of observation, {\it(9)-(11)} emission line ratios, {\it(12)} equivelant width of the $H\alpha$ emission line in \AA, {\it(13)} $\chi^2$ of the STARLIGHT fit, {\it(14)} metallicity, {\it(15)} average age of the stellar population, {\it(16)} age of the most recent starburst event, {\it(17)} classification based on Stasi{\'n}ska et al. (2006), {\it(18)} classification based on the BPT diagrams (Baldwin, Phillips \& Terlevich 1981)}
\newline{{$^\star$} Errors of the [SII] doublet are probably underestimated in the cases that the B telluric band is located near the specific emission lines.}
\end{minipage}
\end{table}
\end{landscape}

\begin{landscape}
\begin{table}
\centering
\begin{minipage}{300mm}
\caption{Observational \& SSP results, emission line ratios and classification.}
\tabcolsep 3pt
\begin{tabular}{lccccccccccccccccc}
{\em NAME} &{No} & {\em RA}       &{\em DEC}     & m &{\em U.T.} & {\em start U.T.} &  {\em exp.}      & {\em $\rm [OIII]/\rm H\beta$}      
&{\em $\rm [NII]/\rm H\alpha$} & {\em $\rm [SII]/\rm H\alpha^\star$}& EW(H$\alpha$)& $\chi^2$ & $<Z>$ &$\rm <logt>$&$\rm logt_{SB}$ 
      & $\rm C_{St}$ & $\rm C_{BPT}$\\
{\em (1)}&{\em (2)}&{\em (3)}&{\em (4)}&{\em (5)}&{\em (6)}&{\em (7)}&{\em (8)}&{\em (9)}&{\em (10)}&{\em (11)}&{\em (12)}&{\em (13)}&{\em (14)}&{\em (15)}&{\em 16)}&{\em (17)}&{\em (18)}\\
\hline
&&&&&&&&\\
ESO 545-G013       &       & 02 24 40.5 & $-$19 08 31 & 14.41                  &\\
             &N1  & 02 24 50.9 & $-$19 08 03 & 16.19   &01/12/06 & 05:13 & 3600 & - & 0.37$\pm0.02$ & 0.37$\pm0.02$  &-26.1$\pm1.2$&1.2&0.038&8.44&6.59&SFN & - \\
NGC 3786           &        & 11 39 42.5 &   +31 54 33 & 13.88                  &\\
      &N1  & 11 39 44.6 &   +31 55 52 & 13.53  &06/03/06 & 06:34 & 3600& 1.08$\pm0.23$ & 0.71$\pm0.08$ & 0.57$\pm0.08$  &-1.5$\pm0.1$&1.0&0.034&9.13&7.44& AGN & TO \\
UGC 12138        &         & 22 40 17.0 &   +08 03 14 & 15.93                  &\\
        &N1   & 22 40 11.0 &   +07 59 59 & 18.77  &08/10/07 & 02:49 & 3600& 4.48$\pm0.19$ & 0.07$\pm0.01$ & 0.18$\pm0.01$  &-42.2$\pm1.5$&1.3&0.041&7.54&6.79&SFN & SFN \\
UGC 7064            &      & 12 04 43.3 &   +31 10 38 & 15.11                  &\\
     &N1B\tablefootmark{2}   & 12 04 45.6 &   +31 11 27 & 16.68 &18/05/07 & 07:11 & 4200& 0.25$\pm0.08$ & 0.38$\pm0.01$ & 0.15$\pm0.01$  &-16.4$\pm0.3$&0.3&0.011&9.66&9.46&SFN & SFN \\
     &N1A  & 12 04 45.2 &   +31 11 33 & 16.68   &SDSS&-&-& 3.55$\pm0.72$& 1.34$\pm0.47$ &
0.83$\pm0.34$ &-1.2$\pm0.3$&0.3&0.021&9.57&9.12& AGN & AGN \\
     &N2   & 12 04 45.1 &   +31 09 34 & 16.33   &06/03/06 & 08:40 & 2100& 0.74$\pm0.13$ & 0.56$\pm0.03$ & 0.30$\pm0.03$  &-15.4$\pm0.6$&2.2&0.012&9.34&6.74& TO & TO \\
IRAS 00160$-$0719    &     & 00 18 35.9 & $-$07 02 56 & 15.73                  &\\
    &N1 & 00 18 33.3 & $-$06 58 54 & 17.80 &06/10/07 &  &  & 0.93$\pm0.06$ & 0.25$\pm0.01$ & 0.44$\pm0.01$  &
-33.1$\pm0.6$&1.3&0.015&9.57&9.35&SFN & SFN \\
ESO 417-G06       &        & 02 56 21.5 & $-$32 11 08 & 15.54                  &\\
       &N1  & 02 56 40.5 & $-$32 11 04 & 17.43   &06/10/07 & 11:08 & 4200& 1.29$\pm0.05$ & 0.21$\pm0.01$ & 0.29$\pm0.01$  &-72.5$\pm1.3$&2.0&0.011&9.19&6.62&SFN & SFN \\
NGC 1241             &     & 03 11 14.6 & $-$08 55 20 & 13.56                  &\\
    &N1   & 03 11 19.3 & $-$08 54 09 & 15.41   &30/11/06 & 08:00 & 3600& 1.02$\pm0.10$ & 0.34$\pm0.01$ & 0.35$\pm0.01$  &-18.2$\pm0.3$&0.7&0.004&9.58&7.14&SFN & SFN \\
NGC 1320        &           & 03 24 48.7 & $-$03 02 32 & 14.59                  &\\
         &N1  & 03 24 48.6 & $-$03 00 56 & 15.07   &25/10/06 & 09:38 & 3600 &-&-&- &- &0.4&0.021&9.64&6.71&ALG & ALG \\
MRK 612          &         & 03 30 40.9 & $-$03 08 16 & 15.78                  &\\
        &N1  & 03 30 42.3 & $-$03 09 49 & 16.13    &29/11/06 & 09:44 & 3600&-&-&- &-&2.2&0.32&7.62&6.97& ALG & ALG \\
NGC 1358     &     & 03 33 39.7 & $-$05 05 22 & 13.98                  &\\
    &N2  & 03 33 23.5 & $-$04 59 55 & 14.95    &21/10/06 & 11:08 & 3600&-&-&- &- &0.9&0.017&9.83&7.26&ALG & ALG \\
NGC 7672          &        & 23 27 31.4 &   +12 23 07 & 15.23                  &\\
       &N1  & 23 27 19.3 &   +12 28 03 & 14.67  &21/10/06 & 05:51 & 3600& - & - & -  & -&1.0&0.035&9.96&7.00&ALG & ALG \\
NGC 7682        &          & 23 29 03.9 &   +03 32 00 & 14.88                  &\\
         &N1   & 23 28 46.6 &   +03 30 41 & 14.64  &25/10/06 & 06:42 & 3600 & 1.25$\pm0.02$ & 0.45$\pm0.01$ & 0.27$\pm0.01$  &-70.6$\pm0.8$&0.9&0.033&8.08&8.23& TO& TO \\
NGC 7743            &      & 23 44 21.1 &   +09 56 03 & 12.16                  &\\
     &N3   & 23 44 05.5 &   +10 03 26 & 16.95   &20/10/06 & 07:00 & 5400& 2.27$\pm0.05$ & 0.07$\pm0.01$ & 0.24$\pm0.01$  &-58.2$\pm1.2$&1.0&0.013&7.94&7.29&SFN & SFN \\
&&&&&\\
\hline
\end{tabular}
\end{minipage}
\tablefoot{{\it(1)-(18)} as in Table 1}
\end{table}
\end{landscape}

\begin{landscape}
\begin{table}
\centering
\begin{minipage}{220mm}
\caption{Basic properties}
\tabcolsep 3pt
\begin{tabular}{lcclcccccccclcclcccccccc}
\multicolumn{12}{l}{\bf Seyfert 1}&\multicolumn{12}{l}{\bf Seyfert2}\\
{\em Name}&{\em No}&{\em T}&{\em D} &{\em $D/D_{AGN}$} &{\em M}& {\em R}&{$I_R$}&  {$I_M$} &{$I_{\rm\Delta M}$}&{\em Sum}&{\em Type}&{\em Name}&{\em No}&{\em T}&{\em D} &{\em $D/D_{AGN}$} &{\em M}& {\em R}&{$I_R$}&  {$I_M$} &{$I_{\rm\Delta M}$}&{\em Sum}&{\em Type}\\
{\em (1)}&{\em (2)}&{\em (3)}&{\em (4)}&{\em (5)}&{\em (6)}&{\em (7)}&{\em (8)}&{\em (9)}&{\em (10)}&{\em (11)}&{\em (12)}&{\em (1)}&{\em (2)}&{\em (3)}&{\em (4)}&{\em (5)}&{\em (6)}&{\em (7)}&{\em (8)}&{\em (9)}&{\em (10)}&{\em (11)}&{\em (12)}\\
\hline
NGC863&N1&-&18.5*&0.27&-16.11&23&1&0&0&1&ALG& ESO545-G013&N1&S&-&-&-18.79&72&0&1&0&1&SFN\\
MRK1400&N1&-&29.0&0.98&-17.34&94&0&0&1&1&SFN   & NGC3786&N1&SABab pec&107.4&0.91&-19.00&13&1&1&1&3&TO\\
NGC1019&N2&-&30.4&0.75&-17.57&76&0&1&1&2&TO    & UGC12138&N1&-&-&-&-15.48&78 &0&0&0&0&SFN\\
NGC1194&N1&-&23.0&0.36&-16.05&31&1&0&0&1&SFN   & IRAS00160-0719&N1&-&11.2&0.41&-15.74&63 &0&0&0&0&SFN\\
NGC1194&N4&SB&35.4&0.55&-17.12&91&0&0&1&1&SFN   & ESO417-G06&N1&-&14.8&0.34&-15.96&56 &0&0&0&0&SFN\\
1H1142-178&N1&-&18.6&0.91&-17.04&11&1&0&1&2&ALG& NGC1241&N1&SBc&32&0.28&-17.54&19 &1&0&0&1&SFN\\
1H1142-178&N2&-&-&-&-16.66&63&0&0&0&0&SFN   & NGC1320&N1&E&42.6&0.47&-17.13&12&1&0&1&2&ALG\\
MRK699&N1&-&21.8*&1.58&-17.49&52&0&1&1&2&TO    & MRK612&N1&-&45.2&1.04&-17.85&28&1&1&1&3&ALG\\
NGC7469&N1&SAcd pec&47.7&0.54&-17.83&19&1&1&1&3&SFN    & NGC1358&N2&S0&71.6&0.70&-17.95&77&0&1&1&2&ALG\\
NGC526A&N1&E&28.0&1.01&-17.97&10&1&1&1&3&AGN   & NGC7672&N1&SA0&76.6&1.98&-18.17&67 &0&1&1&2&ALG\\
NGC526A&N2&SBO/a&87.6&3.15&-18.11&51&0&1&1&2&ALG& NGC7682&N1&SB0 pec&39.6&0.61&-18.96&67 &0&1&1&2&TO\\
NGC526A&N3&-&24.4&0.50&-17.33&59&0&0&0&0&SFN   & NGC7743&N3 &-&-&-&-14.18&43&1&0&0&1&SFN\\
NGC526A&N4&S&13.8&0.88&-17.88&63&0&1&1&2&SFN   & UGC7064 &N2&S&30.2&0.69&-18.30&26&1&1&1&3&TO\\
NGC5548&N1&-&18.4&0.35&-16.66&100&0&0&0&0&SFN   &&&&&&\\
NGC6104&N1&E&27.2&0.69&-18.24&100&0&1&1&2&ALG& &&&&&\\
\hline
\end{tabular}
%\end{minipage}
\tablefoot{{\it(1)} Name of AGN, {\it(2)} number of neighbour, {\it(3)} morphological type , {\it(4)} isophotal diameters at 25.0 B-mag arcsec$^{-2}$ from the RC3 in arcsec (*near-infrared isophotal diameters at 20.0 K-mag arcsec$^{-2}$ from the 2MASS catalogue), {\it(5)} ratio of neighbours to Seyfert diameter , {\it(6)} absolute $O_{maps}$ magnitude, {\it(7)} projected radial separation in $h^{-1}$ kpc, {\it(8)} Index : 0 if R$>50h^{-1}$kpc or 1 elsewise, {\it(9)} Index : 0 if M$>-17.5$ or 1 elsewise, {\it(10)} Index : 0 if $\rm \Delta M>1.5$ or 1 elsewise, {\it(11)} sum of indices 5 to 7, {\it(12)} classification as in Table 1 and Table 2}
\end{minipage}
\end{table}
\end{landscape}

\begin{table*}
\centering
\begin{minipage}{200mm}
\caption{XMM-Newton observations.}
\tabcolsep 3pt
\begin{tabular}{lccccccc}
\label{xmm}
{\em Name} &{\em Neigh. No}&  {\em 2XMM ID} &{\em Opt. Class}&{\em $log L_X$
(\em 0.2-12 keV)}  & {\em Flux (0.2-12 keV)} & {\em X/O offset} & 
{\em HR} \\
      &  &   &     &   {$\rm (erg~s^{-1}$)} & {$\rm
(erg~cm^{-2}~s^{-1}$)}  & {\rm (arcmin)}
&{ }\\
\hline
NGC1194&N1 &   -     & SFN&$<39.59$  &  $<7.3\times 10^{-15}$  & -  & - \\   
NGC1194&N4 &   -     & SFN&$<39.72$ &  $<1.1\times 10^{-14}$  & - & - \\  
NGC526A&N1 &   J012357.0-350410      & AGN&  40.46 & $3.3\times10^{-14}$ & 0.023
& $ -0.28\pm0.09$ \\
NGC526A&N2 &   J012358.1-350653     & ALG&40.75 & $5.9\times 10^{-14}$ &
0.008 & $ 0.05\pm 0.1$ \\
NGC526A&N3 &   -     &SFN& $<$39.65 & $<4.7\times 10^{-15}$ & -  & - \\ 
NGC526A&N4 &  J012359.0-350741      &SFN& 39.95  & $9.4\times 10^{-15}$ &  0.035
& $-0.61\pm 0.29$ \\ 
UGC12138&N1 &   -    &SFN& $<$40.63 & $<2.8\times 10^{-14}$  & - & - \\  
NGC1320&N1 &  J032448.6-030057     &ALG& 39.46 & $1.2\times10^{-14}$ &  0.020
& $ -0.38\pm 0.17$ \\ 
MRK612&N1  &  J033042.5-030949     & ALG&39.59 & $3.4\times 10^{-15}$ & 
0.060 & $-0.67\pm0.24$ \\ 
NGC1358&N2 &  J033323.3-045953     & ALG& 40.19 &  $3.8\times 10^{-14}$ & 
0.044& $ 0.05\pm0.3$ \\ 
NGC7682&N1 &    J232846.7+033041   & TO&  42.04 & $1.30 \times 10^{-12}$ & 
0.026 & $-0.32\pm 0.02$ \\ 
NGC7743&N3  &   -   &SFN&$<$39.44 &  $3.4 \times 10^{-14}$ &  - & - \\
NGC3786&N1 &   J113944.3+315547    &TO& 39.73 & $2.7\times 10^{-14}$ & 0.08 &
$-0.46\pm0.30$ \\ 
\hline
\end{tabular}
\end{minipage}
\end{table*}

\clearpage
\begin{figure*}
\centering
\includegraphics[width=20cm]{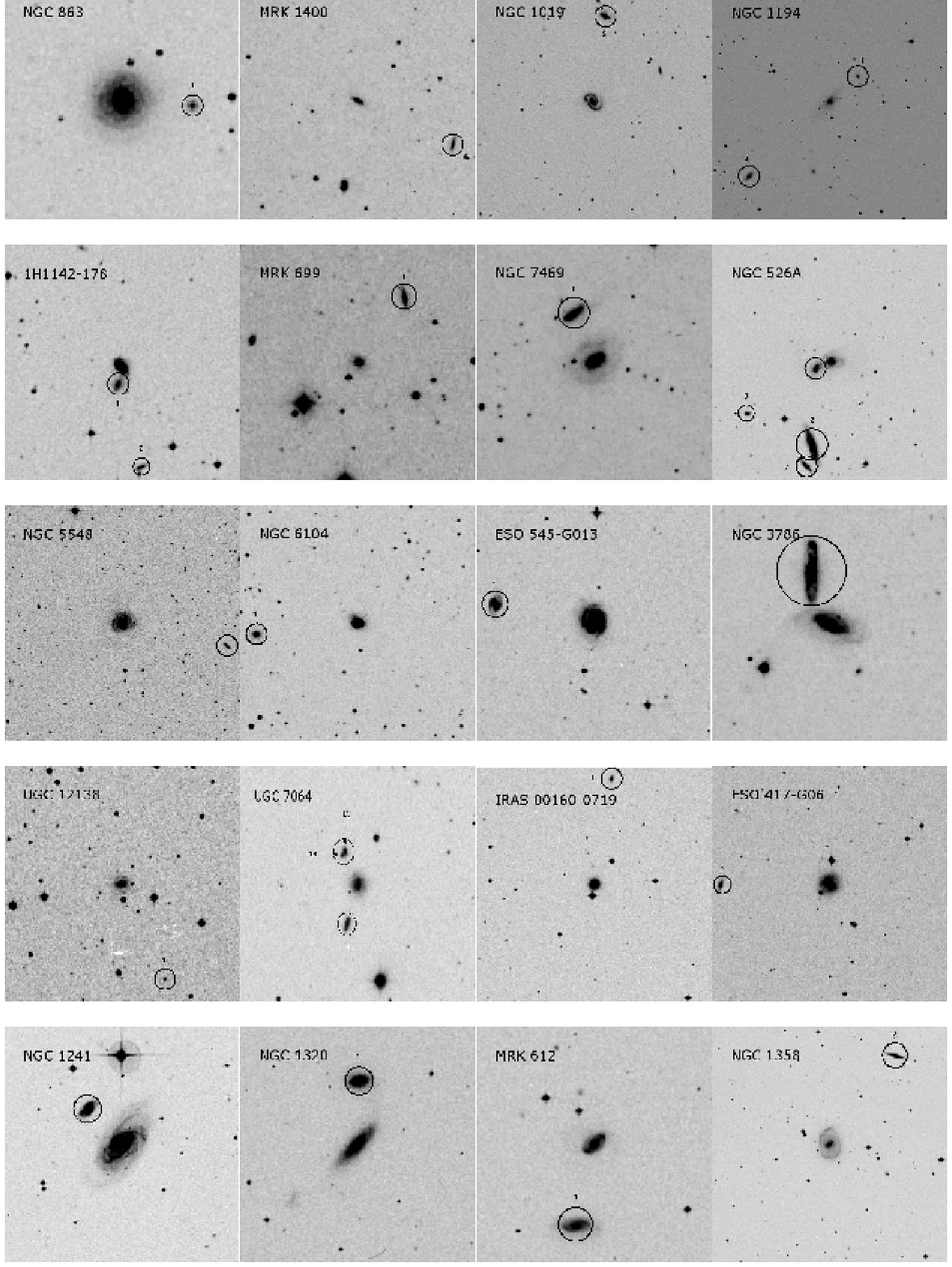}
\end{figure*}

\begin{figure*}
\centering
\includegraphics[width=4cm,height=4cm]{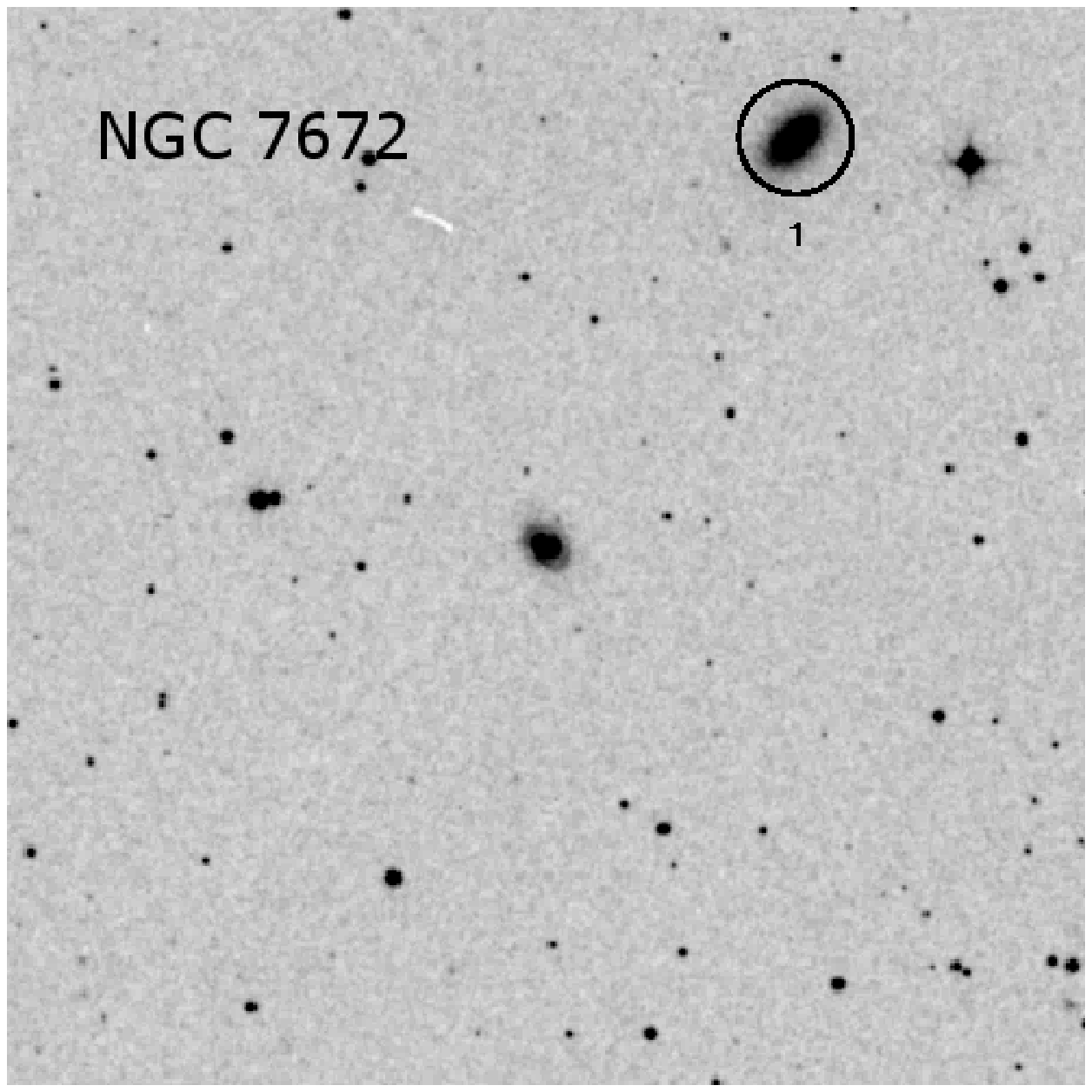}\includegraphics[width=4cm,height=4cm]{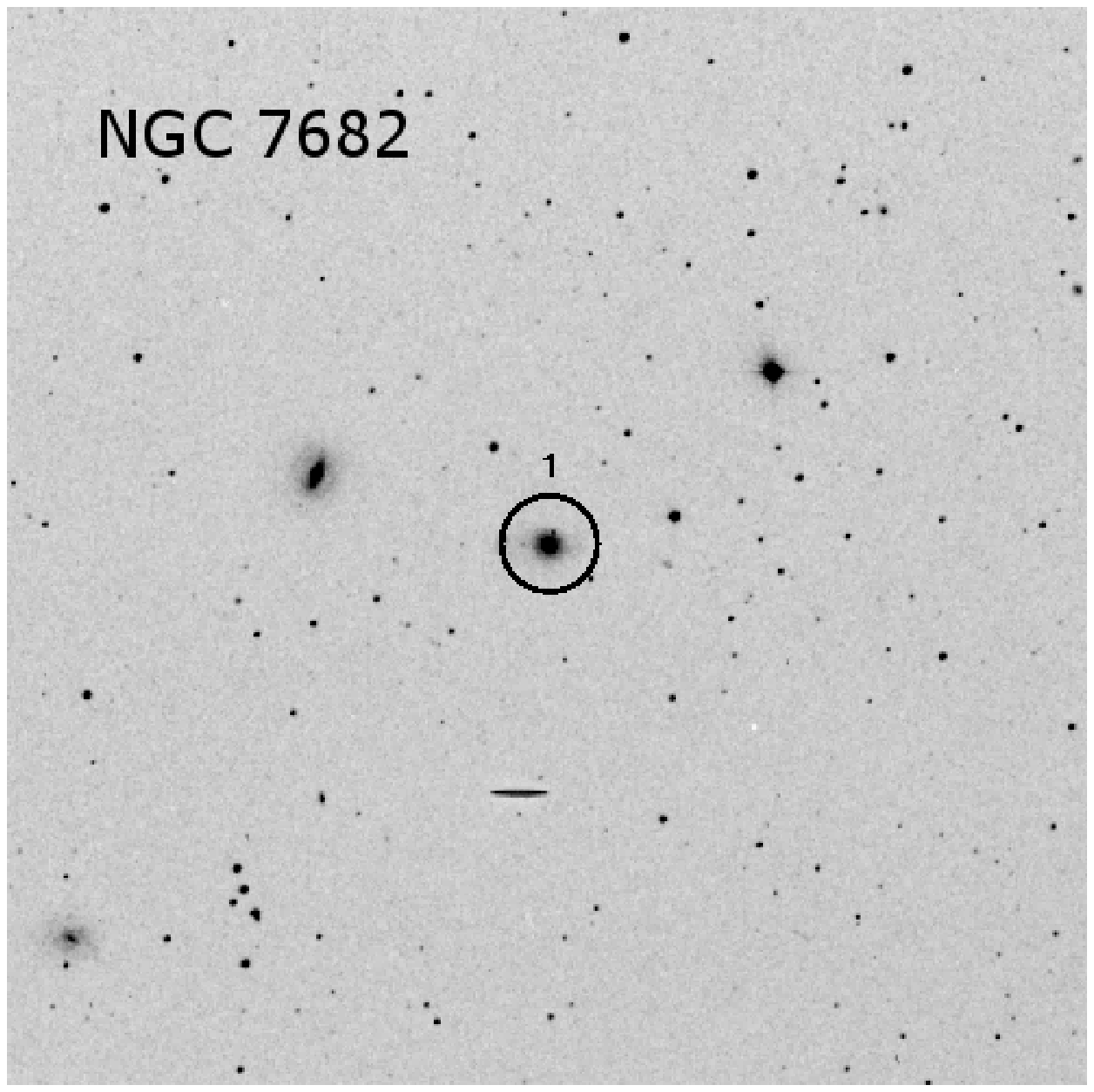}\includegraphics[width=4cm,height=4cm]{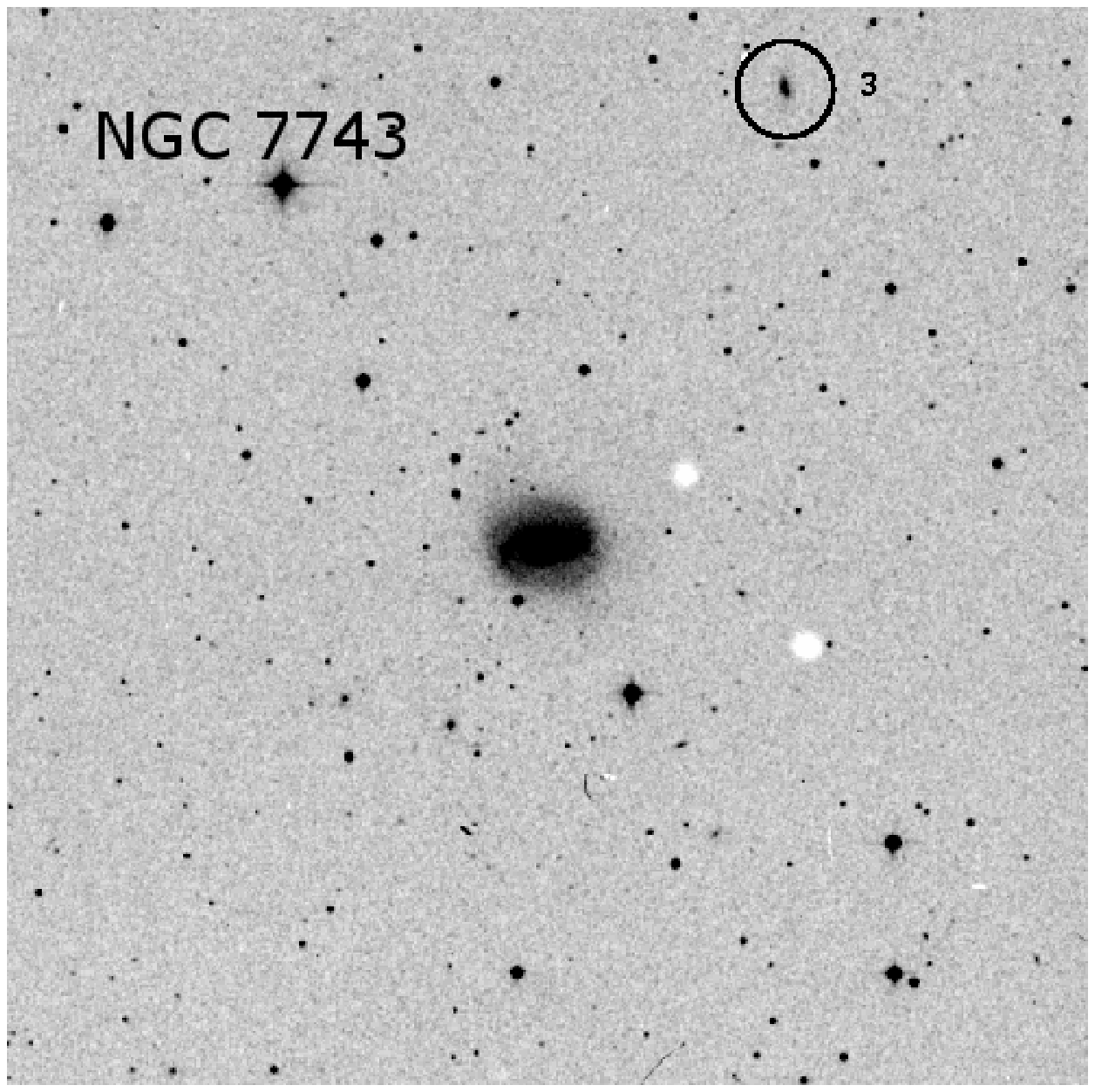}
\caption{Images of the AGN galaxies and their neighbours. The AGN is located in the center of the image except from NGC 7682 which is easily spotted on the left of the image.}
\end{figure*}

\clearpage
%\newpage
\begin{appendix}
\appendixname{ A}
%\end{appendix}
\begin{figure}[h]
\centering
\includegraphics[width=16cm]{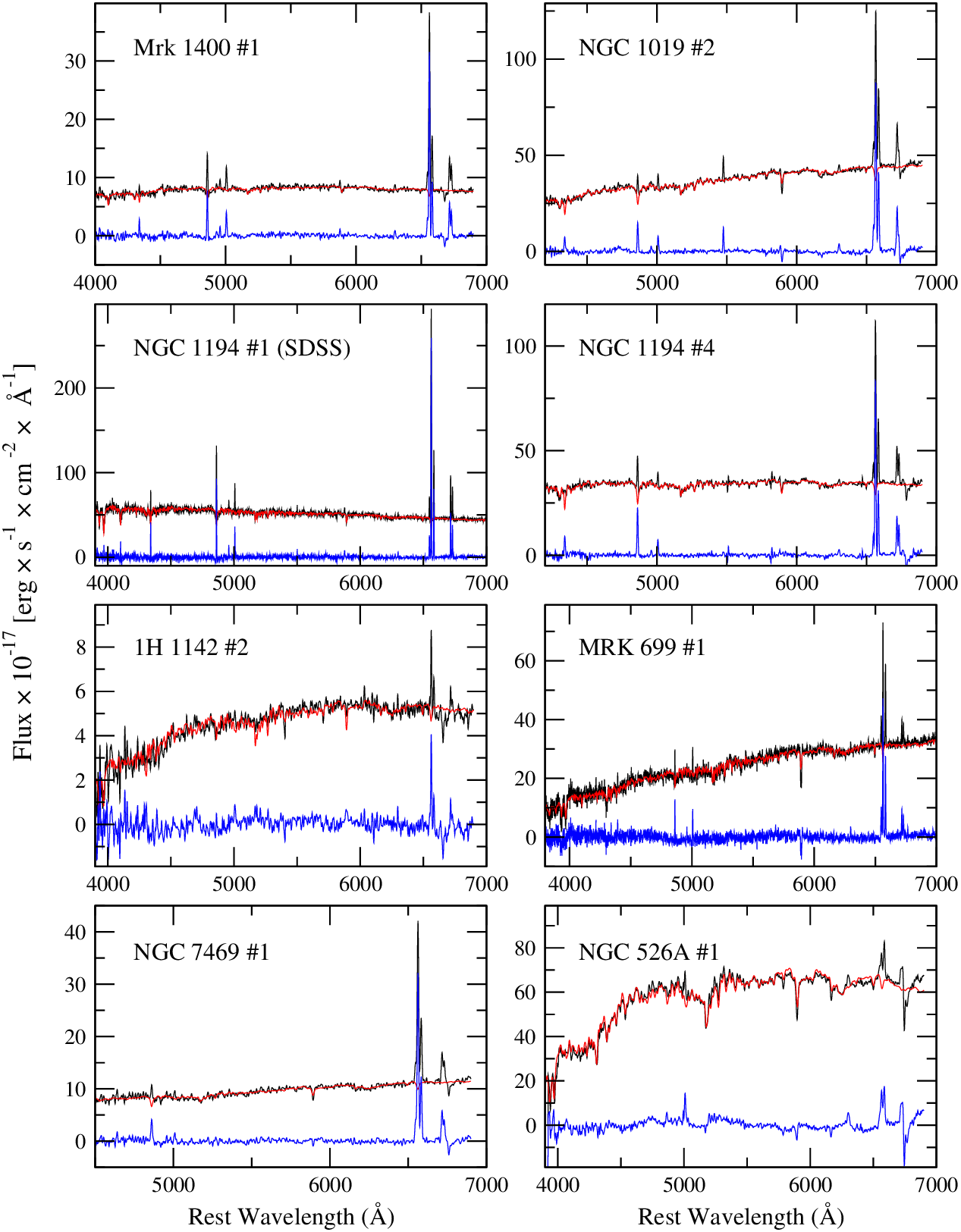}
\end{figure}
\clearpage
\begin{figure*}
\includegraphics[width=16cm]{fig_sp02.eps}
\end{figure*}
\clearpage
\begin{figure*}
\includegraphics[width=16cm,height=16cm]{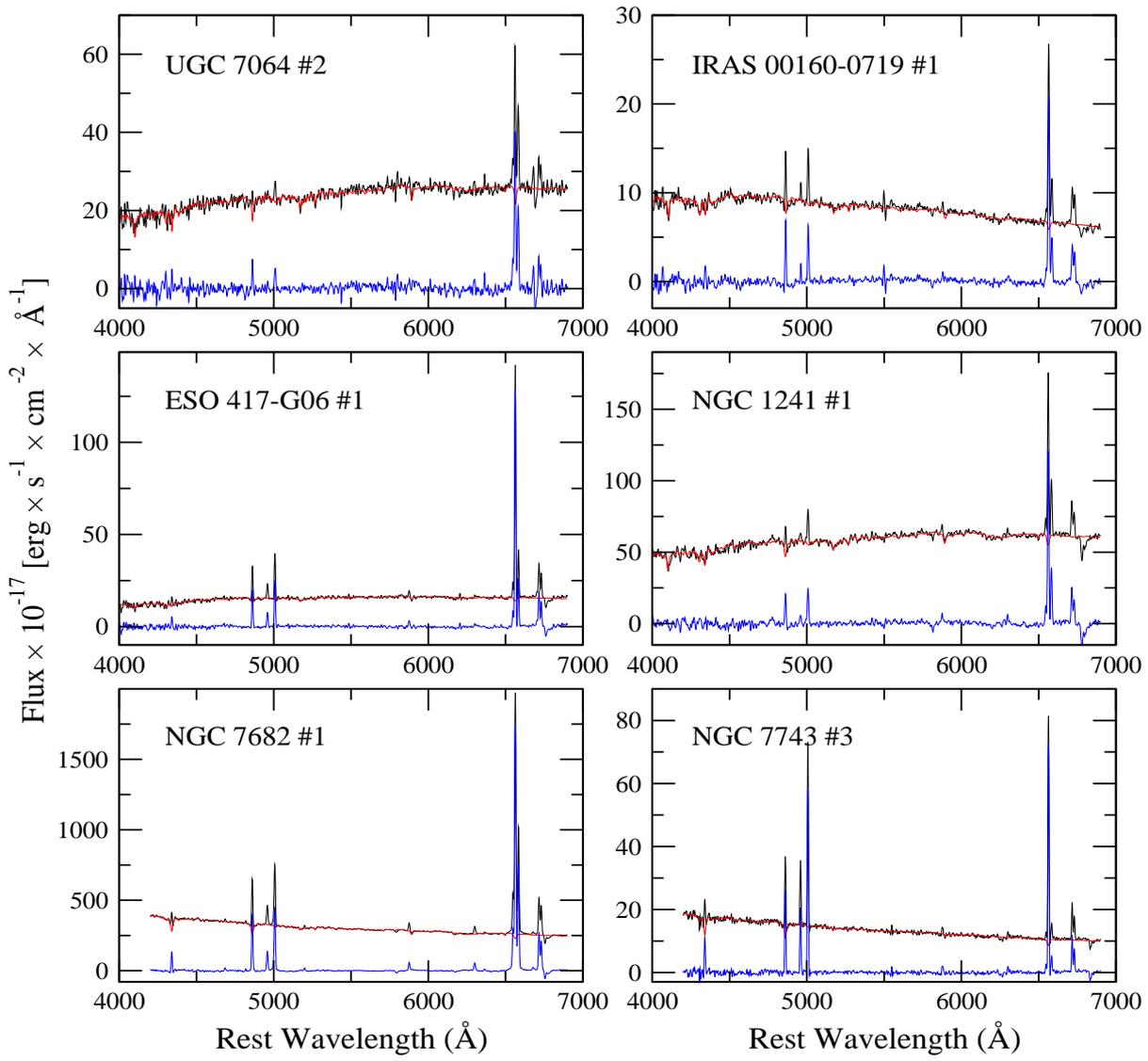}
\caption{Spectra of the ALG neighbours of AGN galaxies, listed in Tables 1 \& 2}
\end{figure*}
\end{appendix}

%\clearpage

%\begin{appendix}
%\noindent
%\appendixname{ B}
%\newline

%The neighbour of MRK 1400 has a typical SFN spectrum, similar to most of the spectra of our samples. Using the GUIAPS package of IRAF we calculate the standard deviation of the flux about the continuum close to the H$\alpha$ and [NII] lines:
%\[\sigma_c=0.9\times10^{-17}\rm erg\, cm^{-2} sec^{-1}\] 
%Given that the EW of the H$\alpha$ line is -37.4 \AA\ and $N_{pix}=4.5$ we derive from equation (1): 
%\[\sigma_{H\alpha}=15.4 \times10^{-17}\rm erg\, cm^{-2} sec^{-1}\] 
%Likewise, for the [NII] line we calculate: 
%\[\sigma_{NII}= 13.1 \times10^{-17}\rm erg\, cm^{-2} sec^{-1}\] 
%Given that the flux of the H$\alpha$ line is $f_{H\alpha}=1154\times10^{-17}$erg\, cm$^{-2}$ sec$^{-1}$ and of the [NII] is $f_{NII}=361\times10^{-17}$erg\, cm$^{-2}$ sec$^{-1}$ we conclude that the 1$\sigma$ standard deviation of the N[II]/H$\alpha$ ratio is $\pm0.01$.
%Using equation (2) we also calculate the standard deviation of the EW which is 0.7 \AA.
%To the above errors we add in quadrature the errors of the Gaussian fitting of the emission lines.
%\end{appendix}

\end{document}